\documentclass{article}
\usepackage{amsmath}

\input{tcilatex}

\begin{document}

\title{Infinite operator-sum representation of density operator for a
dissipative cavity with Kerr medium derived by virtue of entangled state
representation}
\author{Li-yun Hu\thanks{{\small Corresponding author. Address: Department
of Physics, Shanghai Jiao Tong University, Shanghai, 200030, China.
Tel./fax: +86 2162932080.} {\small E-mail addresses: hlyun2008@126.com,
hlyun@sjtu.edu.cn (L.-y. Hu).}} and Hong-yi Fan \\
{\small Department of Physics, Shanghai Jiao Tong University, Shanghai,
200030, China}}
\maketitle

\begin{abstract}
{\small By using the thermo entangled state representation we solve the
master equation for a dissipative cavity with Kerr medium to obtain density
operators' infinite operator-sum representation }$\rho \left( t\right)
=\sum_{m,n,l=0}^{\infty }M_{m,n,l}\rho _{0}\mathcal{M}_{m,n,l}^{\dagger }.$ 
{\small It is noticable that }$M_{m,n,l}${\small \ is not hermite conjugate
to }$\mathcal{M}_{m,n,l}^{\dagger }${\small , nevertheless the normalization 
}$\sum_{m,n,l=0}^{\infty }\mathcal{M}_{nm,,l}^{\dagger }M_{m,n,l}=1${\small %
\ still holds},{\small \ i.e., they are trace-preserving in a general sense.
This example may stimulate further studiing if general superoperator theory
needs modification.}
\end{abstract}

\section{Introduction}

For an open quantum system interacting with the enviroment, one uses
superoperators to describe the evolution of density operator from an initial 
$\rho _{0}$ (pure state or mixed state) into a final\ state $\rho .$ A
superoperator plays a role of linear map from $\rho _{0}\rightarrow \rho ,$
which has an operator-sum representation \cite{r1,r2}%
\begin{equation}
\rho =\sum_{n}M_{n}\rho _{0}M_{n}^{\dagger },  \label{1}
\end{equation}%
where the operators $M_{n}$ and $M_{n}^{\dagger }$ are usually hermite
conjugate to each other, and obey the normalization condition \cite{r5},%
\begin{equation}
\sum_{n}M_{n}^{\dagger }M_{n}=1.  \label{2}
\end{equation}%
$M_{n}$ is named Kraus operator \cite{r3,r4}, $M_{n}$ and $M_{n}^{\dagger }$
are hermite conjugate to each other. Such an operator-sum representation is
the core of POVM (positive operator-valued measure).

In our very recent papers \cite{r6,r7}, based on the thermo entangled state
representation, we have derived some density operators that are in infinite
dimensional operator-sum forms, for example, those for describing
amplitude-damping channel and laser process, the corresponding Kraus
operators are also obtained and their normalization proved, which implies
trace-preserving.

Then an important and interesting question challenges us: is there an
infinite operator-sum representation of density operator in the infinite sum
form $\rho \left( t\right) =\sum_{n=0}^{\infty }M_{n}\rho _{0}\mathcal{M}%
_{n}^{\dagger },$ where the normalization $\sum_{n=0}^{\infty }\mathcal{M}%
_{n}^{\dagger }M_{n}=1$ still holds, but $M_{n}$ and $\mathcal{M}%
_{n}^{\dagger }$ are not hermite conjugate to each other? The answer is
affirmative. In this paper, by using the thermo entangled state
representation we solve the master equation for a dissipative cavity with
Kerr medium to obtain its density operator, and find its infinite
operator-sum representation really possesses such a "strange structure",
this may bring attention of theoreticians who might wonder if the general
theory of superoperators should be modified.

\section{Brief review of the thermo entangled state representation}

In this section, we briefly review the thermo entangled state representation
(TESR). Enlightened by the Thermo Dynamic Theory of Takahashi-Umezawa \cite%
{r8,r9,r10}, we have constructed the TESR in doubled Fock space \cite{r11}, 
\begin{equation}
\left\vert \eta \right\rangle =\exp \left( -\frac{1}{2}|\eta |^{2}+\eta
a^{\dagger }-\eta ^{\ast }\tilde{a}^{\dagger }+a^{\dagger }\tilde{a}%
^{\dagger }\right) \left\vert 0,\tilde{0}\right\rangle ,\;  \label{2.1}
\end{equation}%
in which $\tilde{a}^{\dagger }$ is a fictitious mode accompanying the real
photon creation operator $a^{\dagger },$ $\left\vert 0,\tilde{0}%
\right\rangle =\left\vert 0\right\rangle \left\vert \tilde{0}\right\rangle ,$
$\left\vert \tilde{0}\right\rangle $ is annihilated by $\tilde{a},$ $\left[ 
\tilde{a},\tilde{a}^{\dagger }\right] =1$. Operating $a$ and $\tilde{a}$ on $%
\left\vert \eta \right\rangle $ in Eq.(\ref{2.1}) we obtain the
eigen-equations of $\left\vert \eta \right\rangle $,%
\begin{align}
(a-\tilde{a}^{\dagger })\left\vert \eta \right\rangle & =\eta \left\vert
\eta \right\rangle ,\;\;\;(a^{\dagger }-\tilde{a})\left\vert \eta
\right\rangle =\eta ^{\ast }\left\vert \eta \right\rangle ,  \notag \\
\left\langle \eta \right\vert (a^{\dagger }-\tilde{a})& =\eta ^{\ast
}\left\langle \eta \right\vert ,\ \ \ \left\langle \eta \right\vert (a-%
\tilde{a}^{\dagger })=\eta \left\langle \eta \right\vert .  \label{2.2}
\end{align}%
Note that $\left[ (a-\tilde{a}^{\dagger }),(a^{\dagger }-\tilde{a})\right]
=0,$ thus $\left\vert \eta \right\rangle $ is the common eigenvector of $(a-%
\tilde{a}^{\dagger })$ and $(\tilde{a}-a^{\dagger }).$

Using the normally ordered form of vacuum projector $\left\vert 0,\tilde{0}%
\right\rangle \left\langle 0,\tilde{0}\right\vert =\colon \exp \left(
-a^{\dagger }a-\tilde{a}^{\dagger }\tilde{a}\right) \colon ,$ and the
technique of integration within an ordered product (IWOP) of operators \cite%
{r12,r13}, we can easily prove that $\left\vert \eta \right\rangle $ is
complete and orthonormal,%
\begin{eqnarray}
\int \frac{d^{2}\eta }{\pi }\left\vert \eta \right\rangle \left\langle \eta
\right\vert &=&\int \frac{d^{2}\eta }{\pi }\colon \exp \left( -|\eta
|^{2}+\eta a^{\dagger }-\eta ^{\ast }\tilde{a}^{\dagger }+a^{\dagger }\tilde{%
a}^{\dagger }+\eta ^{\ast }a-\eta \tilde{a}+a\tilde{a}--a^{\dagger }a-\tilde{%
a}^{\dagger }\tilde{a}\right) \colon =1,  \label{2.3} \\
\left\langle \eta ^{\prime }\right. \left\vert \eta \right\rangle &=&\pi
\delta \left( \eta ^{\prime }-\eta \right) \delta \left( \eta ^{\prime \ast
}-\eta ^{\ast }\right) .
\end{eqnarray}%
The $\left\vert \eta =0\right\rangle $ state 
\begin{equation}
\text{ \ }\left\vert \eta =0\right\rangle =e^{a^{\dagger }\tilde{a}^{\dagger
}}\left\vert 0,\tilde{0}\right\rangle =\sum_{n=0}^{\infty }\left\vert n,%
\tilde{n}\right\rangle ,  \label{2.4}
\end{equation}%
possesses the properties%
\begin{eqnarray}
a\left\vert \eta =0\right\rangle &=&\tilde{a}^{\dagger }\left\vert \eta
=0\right\rangle ,\;a^{\dagger }\left\vert \eta =0\right\rangle =\tilde{a}%
\left\vert \eta =0\right\rangle ,\;  \label{2.5} \\
\left( a^{\dagger }a\right) ^{n}\left\vert \eta =0\right\rangle &=&\left( 
\tilde{a}^{\dagger }\tilde{a}\right) ^{n}\left\vert \eta =0\right\rangle .
\end{eqnarray}%
Note that density operators are functions of ($a^{\dagger }$, $a)$, i.e.,
defined in the original Fock space, so they are commutative with operators
of ($\tilde{a}^{\dagger }$, $\tilde{a})$ in the tilde space.

\section{Infinite operator-sum representation of density operator for a
dissipative cavity with Kerr medium}

In the Markov approximation and interaction picture the master equation for
a dissipative cavity with Kerr medium has the form \cite{r14,r15,r15a}%
\begin{equation}
\frac{d\rho }{dt}=-i\chi \left[ \left( a^{\dagger }a\right) ^{2},\rho \right]
+\gamma \left( 2a\rho a^{\dagger }-a^{\dagger }a\rho -\rho a^{\dagger
}a\right) ,  \label{2.6}
\end{equation}%
where $\gamma $ is decaying parameter of the dissipative cavity, $\chi $ is
coupling factor depending on the Kerr medium. Next we will solve the master
equation by virtue of the entangled state representation and present the
infinite sum representation of density operator.

Operating the both sides of Eq.(\ref{2.6}) on the state $\left\vert \eta
=0\right\rangle ,$ letting%
\begin{equation}
\left\vert \rho \right\rangle =\rho \left\vert \eta =0\right\rangle ,
\end{equation}%
and using Eq.(\ref{2.5}) we have the following state vector equation,

\begin{equation}
\frac{d}{dt}\left\vert \rho \right\rangle =\left\{ -i\chi \left[ \left(
a^{\dagger }a\right) ^{2}-\left( \tilde{a}^{\dagger }\tilde{a}\right) ^{2}%
\right] +\gamma \left( 2a\tilde{a}-a^{\dagger }a-\tilde{a}^{\dagger }\tilde{a%
}\right) \right\} \left\vert \rho \right\rangle ,  \label{2.7}
\end{equation}%
the formal solution of $\left\vert \rho \right\rangle $ is%
\begin{equation}
\left\vert \rho \right\rangle =\exp \left\{ -i\chi t\left[ \left( a^{\dagger
}a\right) ^{2}-\left( \tilde{a}^{\dagger }\tilde{a}\right) ^{2}\right]
+\gamma t\left( 2a\tilde{a}-a^{\dagger }a-\tilde{a}^{\dagger }\tilde{a}%
\right) \right\} \left\vert \rho _{0}\right\rangle ,  \label{2.8}
\end{equation}%
where $\left\vert \rho _{0}\right\rangle =\rho _{0}\left\vert \eta
=0\right\rangle ,$ $\rho _{0}$ is the initial density operator. By
introducing the following operators, 
\begin{equation}
K_{0}=a^{\dagger }a-\tilde{a}^{\dagger }\tilde{a},\text{ \ }K_{z}=\frac{%
a^{\dagger }a+\tilde{a}^{\dagger }\tilde{a}+1}{2},\text{ \ }K_{-}=a\tilde{a},
\label{2.9}
\end{equation}%
and noticing $\left[ K_{0},K_{z}\right] =\left[ K_{0},K_{-}\right] =0,$ we
can rewrite Eq.(\ref{2.8}) as%
\begin{align}
\left\vert \rho \right\rangle & =\exp \left\{ -i\chi t\left[ K_{0}(2K_{z}-1)%
\right] +\gamma t\left( 2K_{-}-2K_{z}+1\right) \right\} \left\vert \rho
_{0}\right\rangle  \notag \\
& =\exp \left[ i\chi tK_{0}+\gamma t\right] \exp \left\{ -2t\left( \gamma
+i\chi K_{0}\right) \left[ K_{z}+\frac{-\gamma }{\gamma +i\chi K_{0}}K_{-}%
\right] \right\} \left\vert \rho _{0}\right\rangle .  \label{2.10}
\end{align}%
With the aid of the operator identity \cite{r16}%
\begin{equation}
e^{\lambda \left( A+\sigma B\right) }=e^{\lambda A}\exp \left[ \sigma
B\left( 1-e^{-\lambda \tau }\right) /\tau \right] =\exp \left[ \sigma
B\left( e^{\lambda \tau }-1\right) /\tau \right] e^{\lambda A},  \label{2.11}
\end{equation}%
which is valid for $\left[ A,B\right] =\tau B,$ and noticing $\left[
K_{z},K_{-}\right] =-K_{-},$\ we can reform Eq.(\ref{2.10}) as%
\begin{equation}
\left\vert \rho \right\rangle =\exp \left[ i\chi tK_{0}+\gamma t\right] \exp %
\left[ \Gamma _{z}K_{z}\right] \exp \left[ \Gamma _{-}K_{-}\right]
\left\vert \rho _{0}\right\rangle ,  \label{2.12}
\end{equation}%
where%
\begin{equation}
\Gamma _{z}=-2t\left( \gamma +i\chi K_{0}\right) ,\text{ }\Gamma _{-}=\frac{%
\gamma (1-e^{-2t\left( \gamma +i\chi K_{0}\right) })}{\gamma +i\chi K_{0}}.
\label{2.13}
\end{equation}%
In order to deprive of the state $\left\vert \eta =0\right\rangle $ from Eq.(%
\ref{2.12}), using the completeness relation of Fock state in the enlarged
space $\sum_{m,n=0}^{\infty }\left\vert m,\tilde{n}\right\rangle
\left\langle m,\tilde{n}\right\vert =1$ and noticing $a^{\dagger
l}\left\vert n\right\rangle =\sqrt{\frac{\left( l+n\right) !}{n!}}\left\vert
n+l\right\rangle $, we have%
\begin{align}
\left\vert \rho \right\rangle & =\exp \left[ i\chi tK_{0}+\gamma t\right]
\exp \left[ \Gamma _{z}K_{z}\right] \sum_{l=0}^{\infty }\frac{\Gamma _{-}^{l}%
}{l!}a^{l}\rho _{0}a^{\dagger l}\left\vert \eta =0\right\rangle  \notag \\
& =\exp \left[ i\chi tK_{0}+\gamma t\right] \sum_{l=0}^{\infty }\frac{\Gamma
_{-}^{l}}{l!}\exp \left[ \Gamma _{z}K_{z}\right] \sum_{m,n=0}^{\infty
}\left\vert m,\tilde{n}\right\rangle \left\langle m,\tilde{n}\right\vert
a^{l}\rho _{0}a^{\dagger l}\left\vert \eta =0\right\rangle  \notag \\
& =\sum_{l=0}^{\infty }\frac{\Lambda _{m,n}^{l}}{l!}\exp \left[ -it\chi
\left( m^{2}-n^{2}\right) -t\gamma \left( m+n\right) \right]
\sum_{m,n=0}^{\infty }\left\vert m,\tilde{n}\right\rangle \left\langle m,%
\tilde{n}\right\vert a^{l}\rho _{0}a^{\dagger l}\left\vert \eta
=0\right\rangle ,  \label{2.14}
\end{align}%
where we have set%
\begin{equation}
\Lambda _{m,n}\equiv \frac{\gamma (1-e^{-2t\left( \gamma +i\chi \left(
m-n\right) \right) })}{\gamma +i\chi \left( m-n\right) }.  \label{2.15}
\end{equation}%
Further, using 
\begin{equation}
\left\langle n\right\vert \left. \eta =0\right\rangle =\left\vert \tilde{n}%
\right\rangle ,\text{ }\left\vert m,\tilde{n}\right\rangle =\left\vert
m\right\rangle \left\langle n\right\vert \left. \eta =0\right\rangle ,
\label{2.16}
\end{equation}%
which leads to%
\begin{equation}
\left\langle m,\tilde{n}\right\vert a^{l}\rho _{0}a^{\dagger l}\left\vert
\eta =0\right\rangle =\left\langle m\right\vert \left\langle \tilde{n}%
\right\vert a^{l}\rho _{0}a^{\dagger l}\left\vert \eta =0\right\rangle
=\left\langle m\right\vert a^{l}\rho _{0}a^{\dagger l}\left( \left\langle 
\tilde{n}\right\vert \left. \eta =0\right\rangle \right) =\left\langle
m\right\vert a^{l}\rho _{0}a^{\dagger l}\left\vert n\right\rangle ,
\label{2.17}
\end{equation}%
then Eq.(\ref{2.14}) becomes 
\begin{align}
\left\vert \rho \right\rangle & =\sum_{m,n,l=0}^{\infty }\frac{\Lambda
_{m,n}^{l}}{l!}\exp \left[ -i\chi t\left( m^{2}-n^{2}\right) -\gamma t\left(
m+n\right) \right] \left\vert m,\tilde{n}\right\rangle \left\langle
m\right\vert a^{l}\rho _{0}a^{\dagger l}\left\vert n\right\rangle  \notag \\
& =\sum_{m,n,l=0}^{\infty }\sqrt{\frac{\left( n+l\right) !\left( m+l\right) !%
}{n!m!}}\frac{\Lambda _{m,n}^{l}}{l!}e^{-i\chi t\left( m^{2}-n^{2}\right)
-\gamma t\left( m+n\right) }\left\vert m,\tilde{n}\right\rangle \rho
_{0,m+l,n+l},  \label{2.18}
\end{align}%
where $\rho _{0,m+l,n+l}\equiv \left\langle m+l\right\vert \rho
_{0}\left\vert n+l\right\rangle .$ Using Eq.(\ref{2.16}) again, we see%
\begin{equation}
\left\vert \rho \right\rangle =\sum_{m,n,l=0}^{\infty }\sqrt{\frac{\left(
n+l\right) !\left( m+l\right) !}{n!m!}}\frac{\left( \Lambda _{m,n}\right)
^{l}}{l!}e^{-i\chi t\left( m^{2}-n^{2}\right) -\gamma t\left( m+n\right)
}\rho _{0,m+l,n+l}\left\vert m\right\rangle \left\langle n\right\vert \left.
\eta =0\right\rangle .  \label{2.19}
\end{equation}%
After depriving $\left\vert \eta =0\right\rangle $ from the both sides of
Eq.(\ref{2.19}), the solution of master equation (\ref{2.6}) appears as
infinite operator-sum form%
\begin{eqnarray}
\rho \left( t\right) &=&\sum_{m,n,l=0}^{\infty }\sqrt{\frac{\left(
n+l\right) !\left( m+l\right) !}{n!m!}}\frac{\Lambda _{m,n}^{l}}{l!}%
e^{-i\chi t\left( m^{2}-n^{2}\right) -\gamma t\left( m+n\right) }\left\vert
m\right\rangle \left\langle m+l\right\vert \rho _{0}\left\vert
n+l\right\rangle \left\langle n\right\vert  \notag \\
&=&\sum_{m,n,l=0}^{\infty }\frac{\Lambda _{m,n}^{l}}{l!}e^{-i\chi t\left(
m^{2}-n^{2}\right) -\gamma t\left( m+n\right) }\left\vert m\right\rangle
\left\langle m\right\vert a^{l}\rho _{0}a^{\dagger l}\left\vert
n\right\rangle \left\langle n\right\vert ,  \label{2.20}
\end{eqnarray}%
Note that the factor $\left( m-n\right) $ appears in the denominator of $%
\Lambda _{m,n}$ (see Eq. (\ref{2.15})), (this is originated from the
nonlinear term of $\left( a^{\dagger }a\right) ^{2}$), so moving of all $n-$%
dependent$\ $terms to the right of $a^{l}\rho _{0}a^{\dagger l}$ is
impossible, nevertheless, we can formally express Eq.(\ref{2.20}) as 
\begin{equation}
\rho \left( t\right) =\sum_{m,n,l=0}^{\infty }M_{m,n,l}\rho _{0}\mathcal{M}%
_{m,n,l}^{\dagger },  \label{2.26}
\end{equation}%
where the two operators $M_{m,n,l}$ and $\mathcal{M}_{m,n,l}^{\dagger }$ are
respectively defined as%
\begin{eqnarray}
M_{m,n,l} &\equiv &\sqrt{\frac{\Lambda _{m,n}^{l}}{l!}}e^{-i\chi
tm^{2}-\gamma tm}\left\vert m\right\rangle \left\langle m\right\vert a^{l},%
\text{ }  \notag \\
\mathcal{M}_{m,n,l}^{\dagger } &\equiv &\left\{ \sqrt{\frac{\Lambda
_{n,m}^{l}}{l!}}e^{-i\chi tn^{2}-\gamma tn}\left\vert n\right\rangle
\left\langle n\right\vert a^{l}\right\} ^{\dag },  \label{2.27}
\end{eqnarray}%
to one's regret, $M_{m,n,l}$ is not hermite conjugate to $\mathcal{M}%
_{m,n,l}^{\dagger }$. This example may surprise us to wonder if the general
superoperator form in Eq. (\ref{1}) needs modification.

\section{Further analysis for $\protect\rho \left( t\right) $}

Usng the operator identity $e^{\lambda a^{\dagger }a}=\colon \exp \left[
\left( e^{\lambda }-1\right) a^{\dagger }a\right] \colon $ and the IWOP
technique, we can prove that 
\begin{align}
& \ \sum_{m,n,l=0}^{\infty }\mathcal{M}_{m,n,l}^{\dagger }M_{m,n,l}  \notag
\\
& =\sum_{n,l=0}^{\infty }\frac{\left( n+l\right) !}{n!}\frac{(1-e^{-2t\gamma
})^{l}}{l!}e^{-2n\gamma t}\left\vert n+l\right\rangle \left\langle
n+l\right\vert  \notag \\
& =\sum_{n,l=0}^{\infty }\frac{(1-e^{-2t\gamma })^{l}}{l!}a^{\dag
l}e^{-2\gamma ta^{\dagger }a}\left\vert n\right\rangle \left\langle
n\right\vert a^{l}  \notag \\
& =\sum_{l=0}^{\infty }\frac{(1-e^{-2t\gamma })^{l}}{l!}\colon \exp \left[
\left( e^{-2\gamma t}-1\right) a^{\dagger }a\right] \left( a^{\dagger
}a\right) ^{l}\colon =1,  \label{2.29}
\end{align}%
from which one can see that the normalization still holds, i.e., they are
trace-preserving in a general sense, so $M_{m,n,l}$ and $\mathcal{M}%
_{m,n,l}^{\dagger }$ may be named the generalized Kraus operators.

\section{Reduction of $\protect\rho \left( t\right) $ in some special cases}

In general, Eq. (\ref{2.20}) indicates that the Kerr medium causes phase
diffusion while the field in cavity is in amplitude-damping. In Eq. (\ref%
{2.20}), when the decoherent time $t\rightarrow \infty ,$ the main
contribution comes from the $m=n=0$ term, in this case $\Lambda \rightarrow
1 $, then 
\begin{equation}
\rho \left( t\rightarrow \infty \right) \rightarrow \sum_{l=0}^{\infty
}\left\langle l\right\vert \rho _{0}\left\vert l\right\rangle \left\vert
0\right\rangle \left\langle 0\right\vert =\left\vert 0\right\rangle
\left\langle 0\right\vert ,  \label{2.21}
\end{equation}%
since $\mathtt{Tr}\rho _{0}=1,$ which shows that the quantum system reduces
to the vacuum state after a long decoherence interaction, as expected.

In particular, when $\chi =0$, Eq.(\ref{2.20}) becomes%
\begin{align}
\rho \left( t\right) & =\sum_{m,n,l=0}^{\infty }\sqrt{\frac{\left(
n+l\right) !\left( m+l\right) !}{n!m!}}\frac{(1-e^{-2\gamma t})^{l}}{l!}%
e^{-\gamma t\left( m+n\right) }\rho _{0,m+l,n+l}\left\vert m\right\rangle
\left\langle n\right\vert  \notag \\
& =\sum_{m,n,l=0}^{\infty }\frac{(1-e^{-2\gamma t})^{l}}{l!}e^{-\gamma
ta^{\dagger }a}\left\vert m\right\rangle \left\langle m\right\vert a^{l}\rho
_{0}a^{\dag l}\left\vert n\right\rangle \left\langle n\right\vert e^{-\gamma
ta^{\dagger }a}  \notag \\
& =\sum_{l=0}^{\infty }\frac{(1-e^{-2\gamma t})^{l}}{l!}e^{-\gamma
ta^{\dagger }a}a^{l}\rho _{0}a^{\dag l}e^{-\gamma ta^{\dagger }a},
\label{2.22}
\end{align}%
which corresponds to the amplitude decaying mode, coinciding with the result
in Ref.\cite{r6}. While for $\gamma =0,$ Eq.(\ref{2.20}) reduces to 
\begin{equation}
\rho \left( t\right) =\sum_{m,n=0}^{\infty }e^{-i\chi t\left(
m^{2}-n^{2}\right) }\left\vert m\right\rangle \left\langle m\right\vert \rho
_{0}\left\vert n\right\rangle \left\langle n\right\vert =e^{-i\chi \left(
a^{\dag }a\right) ^{2}t}\rho _{0}e^{i\chi \left( a^{\dag }a\right) ^{2}t},
\label{2.23}
\end{equation}%
which implies a process of phase diffusion, for $\rho _{0}=\left\vert
n\right\rangle \left\langle n\right\vert ,$ then 
\begin{equation}
\rho \left( t\right) =e^{-i\chi \left( a^{\dag }a\right) ^{2}t}\left\vert
n\right\rangle \left\langle n\right\vert e^{i\chi \left( a^{\dag }a\right)
^{2}t}=\left\vert n\right\rangle \left\langle n\right\vert ,  \label{2.24}
\end{equation}%
which shows neither decay nor phase diffusion happens in the Kerr medium.

In summary, we have demonstrated through the above example that the
nonlinear Hamiltonian in master equation may demand us to modify Eq. (\ref{1}%
), the general expression of superoperator.

\textbf{Note added. Recently, we were made aware of Refs.\cite{r17} which
deal with the Kerr medium using thermo field dynamic theory.}

\textbf{Acknowledgement }Work supported by the National Natural Science
Foundation of China under Grant 10775097 and 10874174. \textbf{L.Y. Hu
acknowledges Professor V. Srinivasan for his kind attention about Refs.\cite%
{r17}}.

\end{document}